# Exploring the Properties of the $V_B^-$ Defect in hBN: Optical Spin Polarization, Rabi Oscillations, and Coherent Nuclei Modulation

Irina N. Gracheva[1], Margarita A. Sadovnikova[1], Fadis F. Murzakhanov[1], Georgy V. Mamin[1], Eduard I. Baibekov[1], Evgeniy N. Mokhov[2]

[1] Institute of Physics, Kazan Federal University, Kremlyovskaya 18, Kazan 420008, Russia
[2] Ioffe Institute, Polytekhnicheskaya, 26, St. Petersburg 194021, Russia
E-mail: subirina@gmail.com

**Abstract**

Optically active point defects in semiconductors have received great attention in the field of solid-state quantum technologies. Hexagonal boron nitride, with an ultra-wide band gap $E_g = 6$ eV, containing a negatively charged boron vacancy ($V_B^-$) with unique spin, optical, and coherent properties presents a new two-dimensional platform for the implementation of quantum technologies. This work establishes the value of $V_B^-$ spin polarization under optical pumping with $\lambda_{ext}$ = 532 nm laser using high-frequency ($\nu_{mw}$ = 94 GHz) electron paramagnetic resonance (EPR) spectroscopy. In optimal conditions polarization was found to be $P \approx 38.4$ %. Our study reveals that Rabi oscillations induced on polarized spin states persist for up to 30-40 µs, which is nearly two orders of magnitude longer than what was previously reported. Analysis of the coherent electron-nuclear interaction through the observed electron spin echo envelope modulation (ESEEM) made it possible to detect signals from remote nitrogen and boron nuclei, and to establish a corresponding quadrupole coupling constant Cq = 180 kHz related to nuclear quadrupole moment of $^{14}$N. These results have fundamental importance for understanding spin properties of boron vacancy.

## 1. Introduction

Today, point defects in solid state have acquired particular interest in condensed matter physics through their new opportunities and current unsolved problems [1]. Structural disorders of the crystal lattice, called color centers, can lead to significant changes in the optical and spin properties of the material [2,3]. Comprehensive studies of optically polarized high-spin states ($S \geq 1$) of defects in wide-gap semiconductors, have revealed a new scientific and technical direction for the development of quantum technologies [4,5]. The most studied and widespread defects are the negatively charged NV centers in diamonds with excellent optical and coherent properties that can be detected from single electron spin by ODMR (Optically Detected Magnetic Resonance) methods. [6,7]. Defects with similar properties were found in silicon carbide (SiC), which has fluorescent defects in the infrared range [8]. The interplay between spin, optical and charge states properties of color centers in semiconductors allows to realize a room temperature spintronics [9], quantum sensing [10], quantum information processing with defects [11,12]. Meanwhile, both mentioned bulk crystals (SiC and diamond) are formed by $sp^3$- hybridized atoms of nearly 100% nonmagnetic nuclei ($I = 0$) that reduce the undesirable influence of the lattice atoms on the spin coherence of the color center.

Optically addressable defects with high spin states have recently been discovered in hexagonal Boron Nitride (hBN) and are discussed in detail in the following review articles. [13,14]. hBN is one of the most commonly used van der Waals (vdW) material. It is formed through covalent bonding between boron and nitrogen atoms. These 2D atomic planes are interconnected by weak vdW forces, which enables one to exfoliate thin layers of the materials in a controllable way. Ultra-wide bandgap ($E_g = 6$ eV) of hBN [15] can host a great diversity of deep level impurities or point defects, which give rise to optical transitions and spin properties. The negatively charged boron vacancy with an S=1 triplet spin state is one of the most well-studied and understood color centers in hBN. This defect can be created by electron [16], proton [17], and neutron [18], or ion irradiation of the crystal [19], and possesses photoluminescence in the IR range ($\lambda_{em} \approx 800$ nm). The microscopic model of this defect has been previously established through electron paramagnetic resonance (EPR) [18,20], optically detected magnetic resonance (ODMR) methods [18], and density functional theory (DFT) studies [21,20]. The defect has been identified as a missing boron atom with three equivalent nitrogen nuclei ($^{14}$N $I = 1$) as its nearest neighbors The ground state (GS) of $V_B^-$ center possess a zero-field splitting with $D \approx 3.6$ GHz. The optical excitation of a boron vacancy predominantly populates the spin sublevel in its ground state with $m_S = 0$ through the spin-dependent recombination, as depicted in the inset of Fig. 1 (a). Such induced population inversion manifest itself in a phase change in the EPR spectrum signal. However, the maximum degree of spin polarization and the corresponding dependence on the wavelength, laser source power, and medium temperature are poorly studied

Boron vacancies both at zero magnetic field and under external influence ($\mathbf{B_0} \neq 0$) have particular spin states according to $m_S$ quantum number, that can be manipulated and read out by optical, microwave and radiofrequency means. Thereby the $V_B^-$ in ultra-wide band gap host of hBN combining unique spin and optical properties is considered as new perspective two dimensional platform for quantum technologies i.e. in quantum photonics [22,23], quantum sensing [24–26], and qubits [27–29]. Quantum applications based on spin qubits impose requirements for long coherence times to implement complex operations (spin-to-photon or spin-to-charge conversion), phase stability for quantum sensors and effective optical polarization of triplet spin states. The study of electron-nuclear interactions for boron vacancy has shown that most of the electron spin density (84%) is distributed on the three nearest nitrogen nuclei, which must directly effect on the $V_B^-$ coherent properties [30]. Indeed, the experimental limit of the phase coherence time is equal to $T_2 = 15\ \mu s$ [31] that is close to predicted theoretical value

(18 $\mu s$) [32]. The relaxation mechanisms of $V_B^-$ are spin-spin interaction and diffusion through nuclear spin bath. The development of a boron vacancy as a qubit assumes the implementation of robust Rabi oscillations. The importance of Rabi oscillations underlie in their ability to control the quantum states of qubits, which is necessary for performing quantum operations. Rabi oscillations refer to the resonant absorption and emission of energy between two energy levels of a qubit, which allows the qubit to be controlled by adjusting the frequency, amplitude, and duration of the electromagnetic fields applied to it. Previously, Rabi oscillations were studied for a boron vacancy by the ODMR method at low magnetic fields, the duration of which was within 300 ns – 3 $\mu s$ and only a few oscillation periods were observed [33–37]. The existing time ranges of Rabi oscillations limits the possibilities of performing coherent multi-pulse operations based on a boron vacancy. One of the solutions is to conduct Rabi experiments in high-field (high-frequency – W-band) ranges, in which the wave functions of the triplet center are pure.

In this paper, the optical properties of boron vacancies from the point of view of photon-to-spin conversion are investigated by high-field pulse EPR spectroscopy. The optimal wavelength of laser excitation has been experimentally established with a quantitative estimation of the polarization degree of spin sublevels. We also demonstrate the long term stable Rabi oscillations of the boron vacancy in the high magnetic field. The analysis of coherent nuclear oscillations allowed us to establish the value of the $C_q$ constant of the quadrupole coupling with nitrogen nuclei beyond the first nearest shell.

*1.1 Materials and Methods*

The studied hexagonal boron nitride samples with 1 mm × 1 mm × 0.15 mm dimensions (produced by the HQ Graphene company) were irradiated at room temperature by electrons with $E = 2$ MeV to a total dose $6 \times 10^{18}$ cm$^{-2}$. Pulsed EPR measurements were performed, exploiting the abilities of Bruker Elexsys 680 spectrometer in W-band ($\nu_{MW} = 94$ GHz) range. Electron spin echo (ESE)-detected EPR spectra were recorded using a standard Hahn echo sequence with $\pi = 64$ ns and $\tau = 240$ ns for the W-band. Electron Spin Echo Envelope Modulation (ESEEM) was implemented using a three-pulse sequence: $\pi/2 - \tau - \pi/2 - T - \pi/2$, where the length of $\pi$ was 32 ns, and $\tau$ was changed from 300 ns to 8 ns. The Rabi oscillations were obtained by using a three-pulse sequence $\vartheta_{\text{Rabi}} - T - \pi/2 - \tau - \pi - \tau - ESE$, where the first pulse, conventionally denoted by the symbol $\vartheta$, varies from 4 ns to 7–10 $\mu s$ with a step of 8 ns. The second and third pulses refer to the detecting sequence of the state of magnetization. Laser sources with wavelength from 260 nm to 980 nm were used for optical excitation of the sample through an optical fiber. Low-temperature experiments (200 K – 25 K) were carried out using a helium flow cryostat and an Oxford

## 2. Results and Discussion

*2.1 Optical polarization of boron vacancy.*

$S = 1$ nature of boron vacancy in the hBN crystal with leads to the appearance of a doublet signal on the EPR spectrum (Fig.1). In a strong magnetic field **B₀**, the energy levels of the $V_B^-$ begin to acquire additional energy (the Zeeman term), which removes the degeneracy of magnetic states with $M_S = \pm 1$ forming a triplet spin configuration (Fig. 1a, bottom insert). Meanwhile, the $V_B^-$ initially has the so-called zero-field splitting (ZFS) due to the spin-spin interaction which leads to splitting between energy levels corresponding to $M_S = 0$ and $M_S = \pm 1$ by the value $D = 3.6$ GHz. The sign of the ZFS parameter $D$ as "positive" was established earlier at low-temperature measurements in the high-frequency range (94 GHz) of the EPR spectrometer [18]. Thus, in the spectrum we have the opportunity to observe a low-field component corresponding to the transition between $M_S = +1 \leftrightarrow M_S = 0$, and a high-field component $M_S = 0 \leftrightarrow M_S = -1$ with the selection rules $\Delta M_S = \pm 1$. The following spin Hamiltonian is used to describe the EPR spectrum:

$$H = g\mu_B \mathbf{BS} + D\left(S_z^2 - \frac{S(S+1)}{3}\right) - g_N\mu_N \mathbf{BI} + \sum_{k=1}^{3} A_k I_k S + IPI \qquad (1)$$

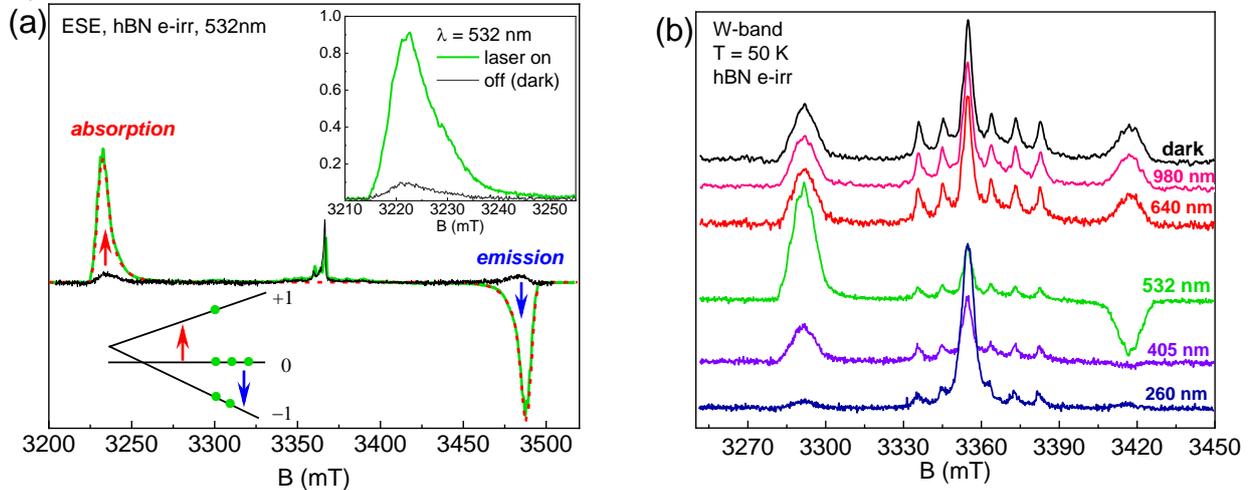

**Figure 1**. (a) The EPR spectrum of the boron vacancy under the laser excitation with a wavelength of λ = 532 nm, where the green solid line – experiment, red dashed line – simulation and the black solid line – dark signal; insert represents detailed spectrum of the low-field component with laser excitation and without (b) EPR spectra dependence on the wavelength λ of laser excitation at the $T = 50$ K of the hBN with magnetic field **B₀** perpendicular to the *c*-axis of the crystal.

Here the first term reflects electron Zeeman interaction with isotropic $g$ - factor, the second term describes ZFS with the principal *z*-axis of the *D*-tensor coinciding with the *c* axis of the crystal. The third term describes the nuclear Zeeman interaction, the fourth term corresponds to the hyperfine interaction of the $V_B^-$ electron spin with the three ($k = 3$) nearest to the vacancy $^{14}$N nuclear spins using the axial *A*-tensor with principal *z*-axis directed along the in-plane nitrogen-dangling bond. The last term is related to the quadrupole interaction, where $P = \frac{3eQ_N V_{ij}}{4I(2I-1)}$ and $C_q = eQ_N V_{zz}$ is the quadrupole coupling constant ($Q_N$ ($^{14}$N) = 0.0193 barn).

Optical activity of the boron vacancy is one of the defect's feature leading to a significant increase of the EPR signal magnitude under laser exposure. The Fig. 1a shows a spectrum with significant increase of signal to noise ratio under laser excitation, where the low-field signal is a classical absorption of microwave energy, while the high-field component is inverted in phase by 180 degrees and represents the induced radiation signal. Laser excitation with a wavelength of λ = 532 nm leads to the predominant population of the non-magnetic state with $M_S = 0$, strongly disturbing the Boltzmann thermodynamic equilibrium. The spin-dependent intercombination conversion mechanism responsible for creating population inversion has been previously identified in other color centers, such as in diamonds [39] and silicon carbide [40].

The quantitative analysis of the optical initialization procedure of the boron vacancy involves the calculation of the spin sublevels polarization degree *P (%)*. Initially, the effect of optical excitation of various laser radiation wavelengths (260 nm – 980 nm) on the intensity of the EPR spectrum of the boron vacancy was investigated (Fig.1b). The highest optical polarization was established for a wavelength of λ = 532 nm. The remaining wavelengths result to only a slight change in the EPR signal magnitudes. The suppression of the EPR signal by ultraviolet radiation (260 nm) may be caused by change of the charge state (-1/0) of the boron vacancy.

In thermodynamic equilibrium and in the absence of optical exposure on the sample, the spin sublevels are populated according to the Boltzmann distribution:

$$n_i = n_0 \cdot e^{-\frac{E_i}{kT}},$$

where $n_0$ is concentration of the centers at energy level $E = 0$, $k$ – Boltzmann constant, $T$ – crystal temperature, $E_i$ - energy level. Under optical excitation $\lambda = 532$ nm, the $|0\rangle$ sublevel becomes the most populated with respect to $|\pm 1\rangle$. The calculation of the population of spin sublevels (see details in Supplementary A) made it possible to estimate the polarization of the dark spectrum ($P_{dark} = 4.5$ %), as well as under the influence of $\lambda = 532$ nm laser exposure ($P_{light} = 38.4$ %).

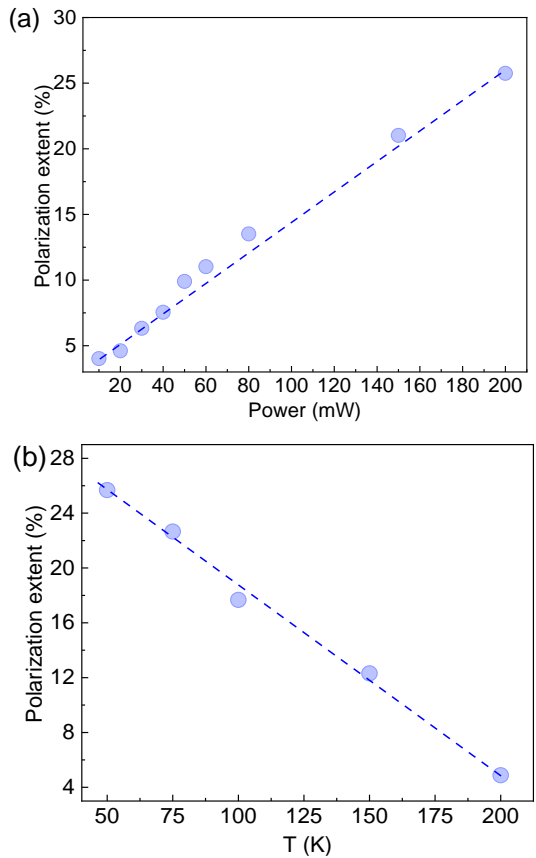

**Figure 2.** Dependence of the polarization degree of the boron vacancy spin states on the (a) laser power at T = 50 K and (b) crystal temperature at P = 200 mW. All measurements were conducted at the perpendicular orientation ($c \perp \mathbf{B_0}$) of the hBN crystal relatively to the magnetic field.

We have investigated the dependence of the polarization degree on the crystal temperature and the power of the laser source. Corresponding dependences are shown in Fig.2a, b. The performed studies reveal that for the boron vacancy in a wide range of laser source power and sample temperature dependence of the polarization degree has a linear character. Such dependence indicates the absence of various types of saturation of the spin system and the possibility of adjusting the spin sublevel population by the power of the optical source with $\lambda = 532$ nm.

## 2.2 Rabi oscillation

The boron vacancy in the hBN crystal is proposed as a potential qubit. In the high-field limit two spin sublevels of $V_B^-$ with quantum numbers $M_S = +1$ and $M_S = 0$ can be considered as a two stable states of the qubit: |0> (ground state) and |1> (exited state), correspondingly. In this case, the Rabi oscillation curves reflect the periodic change of spin population between the ground and the excited states due to a driving external microwave field. In terms of the Bloch sphere the boron vacancy hold an intermediate position between the two spin states forming a superposition and a mixture of two wave functions. Fig. 3 shows the Rabi oscillations of the boron vacancy depending on the microwave power supplied to the resonator system. Coherent oscillations are observed for the $M_Z$ magnetization of the $V_B^-$ electron spin, initially directed along a strong static magnetic field $\mathbf{B_0}$. The oscillation frequency is determined by the magnitude of the $\mathbf{B_1}$ field in the resonator of the EPR spectrometer and the value of the gyromagnetic ratio of the color center: $\nu_{Rabi} = \gamma_e * B_1$, where $\gamma_e = 28$ MHz/T. Consequently, Rabi oscillations lead to periodic spin population changes of qubit states under the stimulated radiation or photons absorption, and the electronic spins hold intermediate positions (superposition). The experiments carried out confirm the linear dependence of the Rabi oscillations frequency on external action strength (microwave exposure) in a wide frequency range (0.5 – 8 MHz) and over several microseconds (up to 30 – 50 µs). The experimental (15.1 µs) [20] and theoretical (18 µs) [41] time limits of phase coherence of the $V_B^-$ center were determined earlier. Here, long Rabi oscillation times have been achieved by high-frequency EPR spectroscopy, exceeding the modulation values (300 ns) detected by ODMR method [33–37].

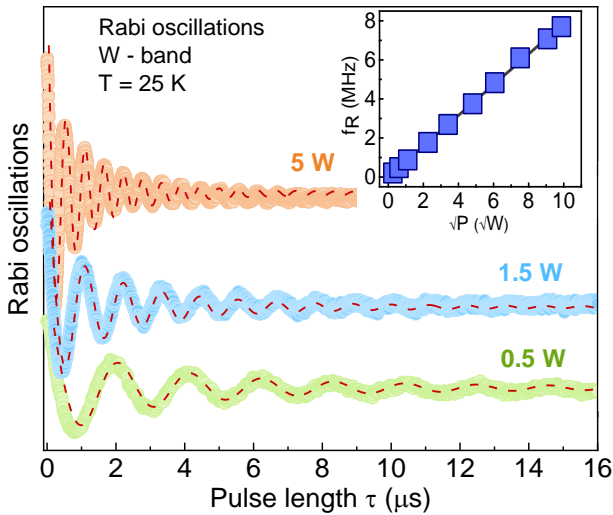

**Figure 3.** Dependence of the boron vacancies Rabi oscillations on the microwave power (5W, 1.5W, 0.5W). Dashed curves show the modeled oscillations using approach described in Supplementary B. The linear relationship between Rabi frequencies and the square root of microwave power is demonstrated in the inset. Dashed curves show the modeled oscillations using approach developed in Supplementary B.

We model our experimental data on electron spin within the approach developed in recent publications (see Supplementary B)

The calculated Rabi oscillation decays agree well with the experimental data accomplished in the studied range of Rabi frequencies 0.2-7.7 MHz (see Fig. 3).

In a crystal, we have a finite number of $V_B^-$ centers that participate in Rabi oscillations. The presence of inhomogeneous local fields and the corresponding distribution of Larmor frequencies contributes to the decay of magnetization and amplitude collapse of oscillations. The collapse time depends inversely on the Rabi oscillation frequency (Fig. 4).

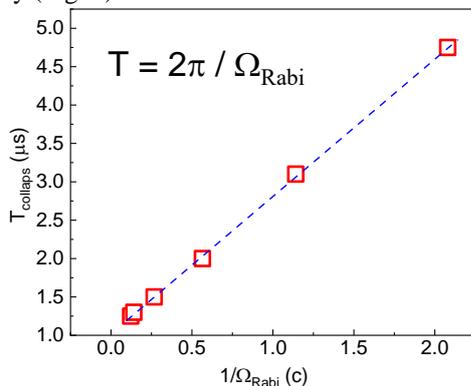

**Figure 4.** Linear dependence of the decay time on the inverse Rabi frequency

It is worth noting that the theoretical simulation was performed using one phase and frequency for each curve, indicating the absence of undesirable two quantum transitions and phase reversals.

### 2.2 Electron spin echo envelop modulation

In the previous section, the influence of nuclear spin diffusion (nuclear spin bath) on the decay of Rabi oscillations was established. Coherent modulations were revealed on the ESE decay curve caused by electron-nuclear interactions of $V_B^-$ spin with surrounding structural magnetic nuclei (Fig. 5, insert) using electron spin echo envelop modulation (ESEEM) spectroscopy pulse sequences. The periodic variation of population between the two electron spin states is induced by driven nuclear sublevel transitions. ESEEM spectrum in the frequency range was obtained by Fourier transform procedure. This technique allow us to detect the signals around the Larmor frequencies of the nitrogen nuclei $^{14}$N ($\nu_L$ = 10.2 MHz) and one of the boron isotopes $^{10}$B ($\nu_L$ = 15.4 MHz). Magnetic nuclei identification is carried out by the calculation of the gyromagnetic ratio ($\gamma_N$) using the nuclear Larmor frequency ($\nu_L$) and the value of the external magnetic field (**B₀**). The corresponding hyperfine interaction values contain anisotropic dipole-dipole term $A_{d-d}$ predicted by ab initio calculations [27,28] and determined by the analysis of EPR spectra [30]. Non-secular matrix elements of hyperfine and quadrupole tensors lead to the possibility of forbidden EPR transitions $m_I = \pm 1$ that are responsible for the occurrence of nuclear oscillations.

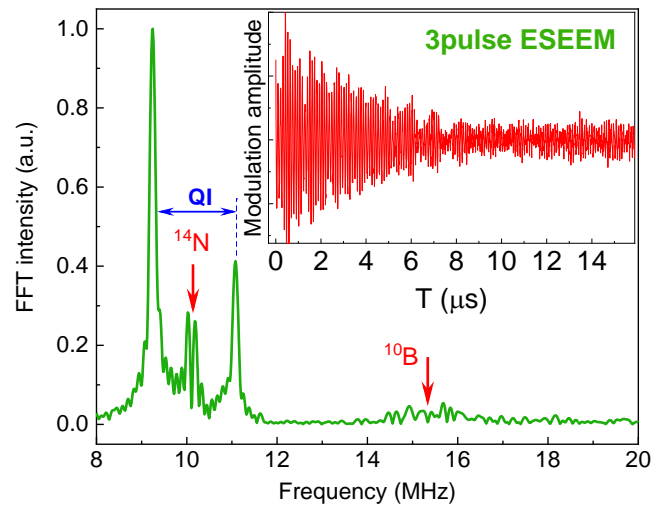

**Figure 5**. Low-field ESEEM spectrum of the hBN sample in the frequency taken at T = 50 K and laser excitation λ = 532 nm. The inset shows the modulation of the electron spin echo due to hyperfine and quadrupole interactions.

The main splitting (QI – quadrupole interaction) for $^{14}$N arises from the interaction of the boron vacancy with three nearest nitrogen nuclei through a quadrupole coupling $C_q$ [16], describing by spin-Hamiltonian (1). The second weak splitting is most likely caused by the interaction of the boron vacancy with distant (remote) nitrogen nuclei. In this case, the value of the quadrupole coupling constant $C_q$ for an undistorted crystal was experimentally determined from Figure 5 $C_q$ = 180 kHz and it is approximately close to predicted value (200 kHz) [42]. Presence of the boron isotope $^{10}$B with a nuclear spin $I$ = 3 leads to unresolved hyperfine and quadrupole structures in the spectra. Detailed description of this structure requires additional studies by electron nuclear double resonance (ENDOR) spectroscopy.

## 3. Conclusions

The optical properties of the boron vacancy have been studied by EPR spectroscopy methods, and the spin polarization $P$ = 38.4% value has been determined under optimal conditions ($T$ = 50 K and $\lambda$ = 532 nm) with the corresponding linear dependence on power of the laser source and the sample temperature. The long-term Rabi oscillations between two spin sublevels in considering the boron vacancy as a qubit were investigated under continues laser irradiation. Analysis of the ESEEM spectroscopy results has shown that the presence of anisotropic hyperfine interactions leading to coherent electron-nuclear modulations of the population of spin sublevels. The value of the quadrupole coupling constant of the boron vacancy with the remote nitrogen nuclei was established $C_q$ = 180 kHz, which is close to the predicted value.

## Acknowledgements

Authors would like to thank Russian Science Foundation grant No 20-72-10068 for its financial support of the W-band EPR/ESEEM experiments, electron irradiation, and interpretation of the experimental data. E. N. Mokhov acknowledges support of state assignments of the Ministry of Science and Higher Education of the Russian Federation to Ioffe Institute (0040-2019-0016) for primary characterization of the samples prior irradiation treatments.